\documentclass{ws-ijmpd}
\usepackage{xcolor} % added by user
\usepackage[super,compress]{cite}
\begin{document}

\markboth{G. Wang, D. Gao and W.-T. Ni et al}
{Orbit Design for Space Atom-Interferometer AIGSO}

\catchline{}{}{}{}{}
%
%%%%%%%%%%%%%%%%%%%%%%%%%%%%%%%%%%%%%%%%%%%%%%%%%%%%%%%%%%%%%%%%%%%%

\title{Orbit Design for Space Atom-Interferometer AIGSO}

\author{Gang Wang$^\ast$, Dongfeng Gao$^{\ast,\dagger}$, Wei-Tou Ni$^{\ast,\ddagger}$, Jin Wang$^{\ast,\dagger}$ and Mingsheng Zhan$^{\ast,\dagger}$}

\address{$^\ast$State Key Laboratory of Magnetic Resonance and Atomic and Molecular Physics, \\
Wuhan Institute of Physics and Mathematics, Chinese Academy of Sciences, Wuhan 430071, China \\
 $^\dagger$Center for Cold Atom Physics, Chinese Academy of Sciences, Wuhan 430071, China \\
 $^\ddagger$National Astronomical Observatories, Chinese Academy of Sciences, Beijing 100012, China\\
 gwanggw@gmail.com\\ dfgao@wipm.ac.cn\\ wei-tou.ni@wipm.ac.cn \\ mszhan@wipm.ac.cn
}

\maketitle

\begin{history}
\received{Day Month Year}
\revised{Day Month Year}
\end{history}

\begin{abstract}
Atom Interferometric Gravitational-wave (GW) Space Observatory (AIGSO) is a mission concept mainly aimed at the middle-frequency (0.1 Hz - 10 Hz) GW detection. AIGSO proposes to have three spacecraft in linear formation with extension of 10 km. The three spacecraft need to maintain 5 km + 5 km constant arm-length formation. In this study, we address the issue of orbit design and thruster requirement. The acceleration to maintain the formation can be designed to be less than 30 pm/s$^2$ and the thruster requirement is in the 30 nN range. Application to other arm-length-maintaining missions is also discussed.

\end{abstract}


\ccode{PACS numbers: 04.80.Nn, 04.80.-y, 07.60.Ly, 95.10.Eg, 95.55.Ym}

%\tableofcontents

\section{Introduction}

After the direct detection of high-frequency gravitational waves (GWs) from binary black holes\cite{LVC2016} and from binary neutron stars\cite{LVC2017}, activities on the middle-frequency GW detection have been significantly increased. The middle-frequency GW band classification (0.1 Hz - 10 Hz) comes from it is in between the Earth-based high-frequency detectors and the space-borne low-frequency detectors (Doppler tracking of spacecraft\cite{Armstrong2003}, LISA\cite{LISA2000} and ASTROD\cite{ni2008,ASTROD}). The first middle-frequency laser-interferometric space GW detection concepts are DECIGO\cite{Kawamura2006} and BBO\cite{Crowder&Cornish}. The new low-frequency laser-interferometric space GW projects include TAIJI\cite{Hu&Wu} and Tianqin\cite{Luo2015}. The new middle-frequency laser-interferometric space GW projects include AMIGO\cite{ni&wang&wu,ni2018} and B-DECIGO\cite{B-DECIGO,Kawamura2018}. The middle-frequency GW proposals also include torsion-bar proposal TOBA,\cite{TOBA,TOBA2} superconducting sensing proposal SOGRO\cite{SOGRO,SOGRO2} and Doppler tracking proposal using optical clocks INO\cite{INO}. Several groups have made GW detection proposals\cite{borde2004,chiao2004,roura2006,foffa2006,delva2006,tino2007,kasevich2008,hogen2011,gao2011,kasevich2016,graham2017,chaibi2016,canuel2018,gao2018,zaiga2018,Graham&Jung2018} using atom interferometers (AI) after AIs have reached high precision. Earth-based proposals include Tino's \cite{tino2007}, AGIS \cite{kasevich2008}, MIGA \cite{chaibi2016,canuel2018}, and ZAIGA\cite{zaiga2018}. Space-borne proposals include Stanford's proposals \cite{hogen2011,kasevich2016,graham2017} and AIGSO\cite{gao2018}. 

In Ref. \refcite{gao2018}, we have studied a new middle-frequency GW detection proposal, called the atom interferometric gravitational-wave space observatory (AIGSO). The proposal uses three drag-free satellites, with arm length of 10 km. Compared to Standford's proposals\cite{hogen2011,kasevich2016,graham2017}, AIGSO uses atomic matter waves both as spacecraft links and interferometry readout. Compared to space-borne laser interferometric GW detectors, our scheme is much smaller in size. In this paper, we will focus on the issue of orbit design for AIGSO. The rest of the paper is organized as follows. In Section 2, we give an overall description of AIGSO. In Section 3, we calculate the geodesic mission orbit. In Section 4, orbit correction and thruster requirement are discussed. Comments and conclusions are given in Section 5.

\section{AIGSO}  \label{sec:AIGSO}

\subsection{AIGSO configuration} \label{sec:2.1}

\begin{figure}[htb]
   \centering
   \includegraphics[width=0.7\textwidth]{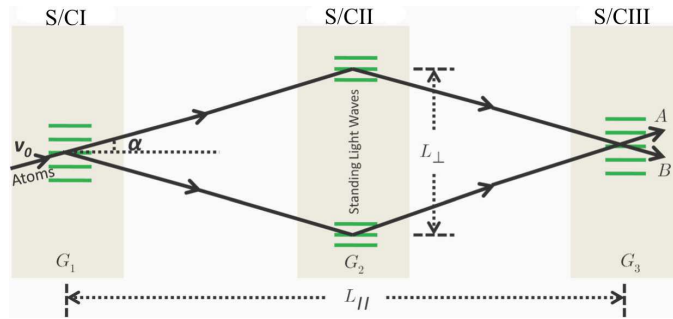}
   \caption{Schematic diagram of the AIGSO from Ref. \cite{gao2018}.}
   \label{fig:AIGSO_SCs}
\end{figure}

A schematic diagram of AIGSO is given in Fig. \ref{fig:AIGSO_SCs}, where the length and width are denoted by $L_{//}$ and $L_{\perp}$, respectively. The atomic source and the first standing light wave are produced in Satellite 1. The middle two standing light waves are hosted in Satellite 2. Satellite 3 is used to house the final standing light wave and the atom detection terminals A and B. 

Inside S/CI, the supersonic atomic beam is first produced with initial velocity $v_0$. Then, the atomic beam is equally split into two beams by the first standing light wave. After a propagation time of $\frac{T}{2} \equiv \frac{L_{//}}{2 v_0 {\rm cos} \alpha}$ , the two atomic beams reach S/CII, and are reflected by the middle two standing light waves. With another propagation time $\frac{T}{2}$, the two beams are recombined to complete the final interference. Complementary interference fringes can be formed in terminals A and B, where the GW induced signals can be read out.

The AIGSO proposes to have 3 S/Cs in a straight linear formation of 10 km ($L_{//}$) extension as shown in Fig. \ref{fig:AIGSO_SCs}. 
In our orbit calculation, we concern the two end S/C orbits, one is the atom sending satellite (S/CI in Fig. \ref{fig:AIGSO_SCs}), another is atom receiving satellite (S/CIII in Fig. \ref{fig:AIGSO_SCs}). Once these two S/Cs orbits are settled, the orbit of  the middle satellite (Satellite II in Fig. \ref{fig:AIGSO_SCs}) could be solved by interpolation. Hence, in the orbit design we need only consider the two end S/Cs.

Suppose the incoming $h_\times$-polarization GWs propagate along the direction orthogonal to the interferometer. Take $h_\times =h \, {\rm e}^{i (2\pi f t + \phi_0)}$, where $h$ and $f$ are the amplitude and frequency of the GW, respectively. $\phi_0$ is the initial phase of the GW. The GW-induced phase shift is calculated to be
\begin{equation}
\Delta\Phi=\frac{N k_l h L_{//}}{2} \tilde{C}(f){\rm e}^{i \phi_0},
\label{phase}
\end{equation}
where
\begin{eqnarray}
\tilde{C}(f)&=&\frac{1}{2} \left(\frac{\pi f T}{2} \sin(\pi f T) + \frac{\cos(\pi f T)- \cos(2\pi f T)}{2}-\frac{\sin(2\pi f T)- \sin(\pi f T)}{\pi f T}\right) \nonumber\\
&&+ \frac{i}{2} \left(-\frac{\pi T f }{2} \cos(\pi f T) +\frac{1+\cos(2\pi f T)- 2\cos(\pi f T)}{\pi f T}-\frac{\sin(2\pi f T)- \sin(\pi f T)}{2}\right).
\end{eqnarray}
It is clear that Eq. (\ref{phase}) only depends on the parameters: $L_{//}$, $T$, $k_l$, and $N$. $k_l=2\pi/\lambda_l$ is the wavenumber of the standing light waves, and $N$ is the number of photon momentum transfer to the atom. According to Ref. \cite{gao2018}, we will take Argon as our example. The size of AIGSO is proposed to be $L_{//} = 10^4 \, {\rm m}$. The interrogation time is T=10 s. To split the Ar beam by angle $\alpha=10^{-4}\, {\rm rad}$, we need $N=10$ and $\lambda_l=810$ nm.

Shot noise is the fundamental limit on the detection sensitivity of any interferometric GW detector. For our AIGSO, the shot-noise-limited sensitivity, $\tilde{h}_{sh} (f)$, can be easily calculated from Eq. (\ref{phase}),
\begin{eqnarray}
 \tilde{h}_{sh} (f)=\frac{\lambda_l }{\pi N L_{//} |\tilde{C}(f)|}\frac{1}{\sqrt{\mathcal{R}}}, 
\end{eqnarray}
where $\mathcal{R}$ is the flux intensity of the atomic beam, which is proposed to be $\mathcal{R}\sim {\rm 10^{16} \, atoms/s}$.

\begin{figure}[htb]
   \centering
   \includegraphics[width=0.85\textwidth]{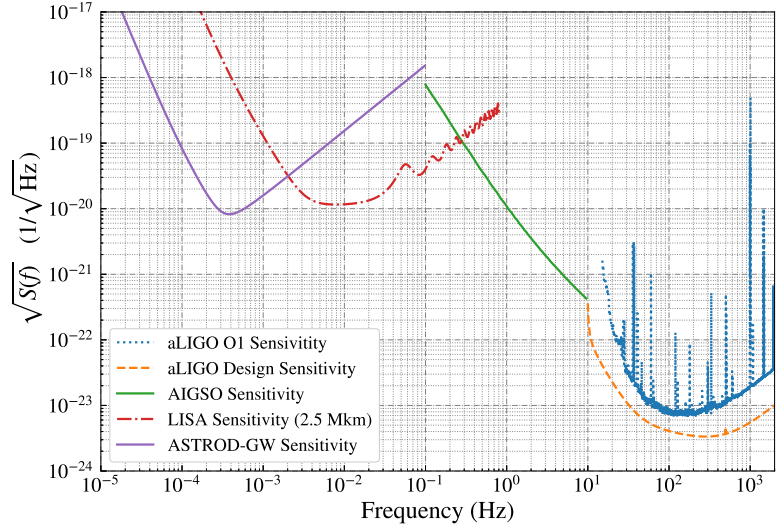}
   \caption{The sensitivity curves of advanced LIGO\cite{LVC2016,barsotti2018}, AIGSO, LISA\cite{cornish2018} and ASTROD-GW\cite{ni2008}.}
   \label{fig:multi_sensitivity}
\end{figure}

The amplitude spectral density for the acceleration noise is
\begin{eqnarray}
 \tilde{h}_{ac} (f)=\frac{\sqrt{S_{ac}}}{(2\pi f)^2 L_{//}}.
\end{eqnarray}
Supposing $S^{1/2}_{ac}$ is about $3 \times 10^{-15}\, {\rm m \cdot s^{-2}/ \sqrt{Hz}}$, which is the LISA requirement\cite{LISA2000}, the sensitivity curve of AIGSO with the advanced LIGO (aLIGO), LISA and ASTROD-GW are shown in Fig. \ref{fig:multi_sensitivity}.

\subsection{Orbit Design} \label{sec:orbit-design}

The purpose of mission orbit design for AIGSO is to have the arm lengths among three S/Cs satisfying the stability requirement.
There are various methods to design the orbit which could be rotating around the Earth or Sun. In reference \refcite{Graham&Jung2018}, a single-baseline AI configuration with 2 S/Cs separated by $2 \times 10^4$ km orbit around the Earth was studied for GW sensitivity. 
In this work, we demonstrate the feasibility of a single-baseline 3 S/Cs formation in an Earth-like solar orbit AIGSO. To achieve the targeting orbit, the Clohessy and Wiltshire frame (CW frame) for one body problem could be used to preliminarily design the orbit to keep the approximately constant distance between S/Cs\cite{Dhurandhar+etal+2005,wang&ni2012,ni2016}.
In our past works, we used CW frame to calculate/simulate the constellations orbits for LISA\cite{wang&ni2013CQG}, TAIJI\cite{wang&ni2017}, and AMIGO\cite{ni&wang&wu} for obtaining preliminary initial conditions and then using the numerical method to optimize them. Now we apply the same procedure to the AIGSO mission orbit design.
\begin{figure}[hbt]
   \centering
   \includegraphics[width=0.7\textwidth]{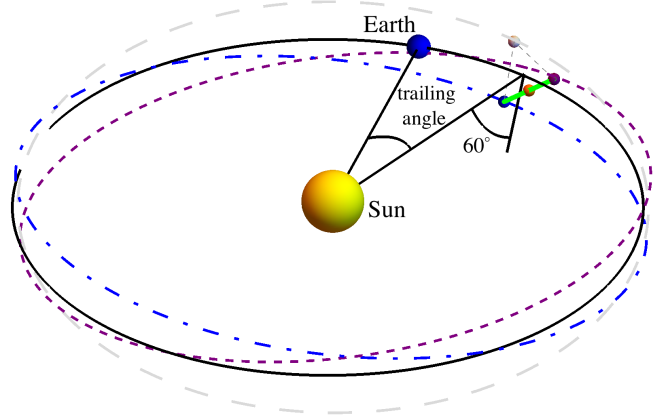}
   \caption{Schematic orbit of the AIGSO around the Sun.}
   \label{fig:AIGSO_arm}
\end{figure}

We explore two Earth-trailing initial configurations for the AIGSO with trailing angle 10 degrees and 2 degrees. As we can expect, for the larger trailing angle configuration, the mission orbit could have more stable arm length because of the larger distance from the Earth gravitational perturbation. However, it may require larger distance communication, better control system, and a higher budget etc. The smaller trailing angle configuration would suffer more gravitational perturbation and would have shorter time in stabler mission orbit configuration.

Just like AMIGO, we initially calculate the AIGSO mission orbit as a LISA-like configuration but with three 10 km arms. AIGSO proposes to arrange three S/Cs in straight 10 km baseline as reviewed in Section \ref{sec:2.1}. Then we choose the most stable arm from the three candidate arms as the configuration of AIGSO. Same as LISA-like orbit, the orbital plane formed by S/Cs will have nominal 60$^\circ$ inclination w.r.t the ecliptic plane as shown in Fig. \ref{fig:AIGSO_arm}. The S/Cs follow the geodesic and the length of baseline fluctuates with time. These orbits are geodesic mission orbits which are studied in Section \ref{sec:geodesic}.

For AIGSO mission, it is essential to have a fixed baseline. In this case, at least two of three S/Cs will not follow the geodesics. From the achieved geodesic orbit, we can choose one free-fall S/C as a benchmark. Other two S/Cs follow it and adjust their position using thrusters to form a steady baseline. This configuration is a steady mission configuration. We simulate it and study the basic thruster requirement in Section \ref{sec:fixed-baseline}.

\section{Geodesic Mission Orbit} \label{sec:geodesic}

\subsection{Orbit Selection}

For a LISA-like orbit configuration, the orbit of each S/C is elliptic having an eccentricity $e$ and inclination $\iota$ as described in Section \ref{sec:orbit-design}. The angular velocity of frame rotation is $\Omega$ as same of the orbit rotation.
Following the algorithm in reference \refcite{Dhurandhar+etal+2005} and our previous work in references \refcite{wang&ni2013CQG,wang&ni2017,wang2011,wang&ni2013CPB,dnw2013,wang&ni2015}, the first order parameter is given by $\alpha [ = \iota = l/(2R)]$, where $l=10$ km is the nominal arm length between the S/Cs and the $R$ is the radius of the S/Cs orbit. We initially choose the time $t_0 =$ JD2462503.0 (2030-Jan-1st 12:00:00) as the starting point for the science observation. A set of S/C initial conditions in the heliocentric elliptical coordinate is defined in reference \refcite{Dhurandhar+etal+2005},
 \begin{equation} \label{equ:1}
 \begin{split}
   X_k & = R ( \cos \psi_k + e) \cos \epsilon \\
   Y_k & = R \sqrt{ 1 - e^2} \sin \psi_k \qquad (k = 1, 2, 3), \\
   Z_k & = R( \cos \psi_k + \epsilon ) \sin \epsilon
 \end{split}
\end{equation}
where $\epsilon \simeq 3.34 \times 10^{-8} $; orbital eccentricity $e \simeq 1.93 \times 10^{-8}$;  $R = 1$ AU; $\psi_k$ is the eccentric anomaly which is related to the mean anomaly $ \Omega (t - t_0)$, and $\Omega$ is the average angular velocity which is 2$\pi$/(one sidereal year). The $\psi_k$ could be obtained by solving the equation numerically
 \begin{equation}
   \psi_k + e \sin \psi_k = \Omega (t - t_0) - (k-1) \frac{2\pi}{3}.
 \end{equation}
 The $x_k , y_k, z_k (k=1,2,3)$ are defined as
 \begin{equation}
 \begin{split}
   x_k & = X_k \cos \left[ \frac{2\pi}{3} (k-1) + \varphi_0  \right] - Y_k \sin \left[\frac{2\pi}{3} (k-1) + \varphi_0 \right] \\
   y_k & = X_k \cos \left[ \frac{2\pi}{3} (k-1) + \varphi_0 \right] + Y_k \sin \left[\frac{2\pi}{3} (k-1) + \varphi_0 \right], \\
   z_k & = Z_k
 \end{split}
\end{equation}
where $\varphi_0 = \psi_E - \theta$ and $\psi_E$ is the position angle of Earth w.r.t. the X-axis at $t_0$ in the ecliptic plane.
Then the initial positions of the S/Cs in the heliocentric coordinate are
\begin{equation} \label{equ:initialcondition}
 \begin{split}
  \mathbf{r}_{\mathrm{S/Ck}} &= [x_k , y_k , z_k] \quad (k = 1, 2, 3). \\
 \end{split}
\end{equation}
The initial velocities are obtained by calculating the derivatives of Eq. (\ref{equ:1}) at $t = t_0$. 

\subsection{Numerical Mission Orbits}

\begin{figure}[htb]
   \centering
   \includegraphics[width=0.49\textwidth]{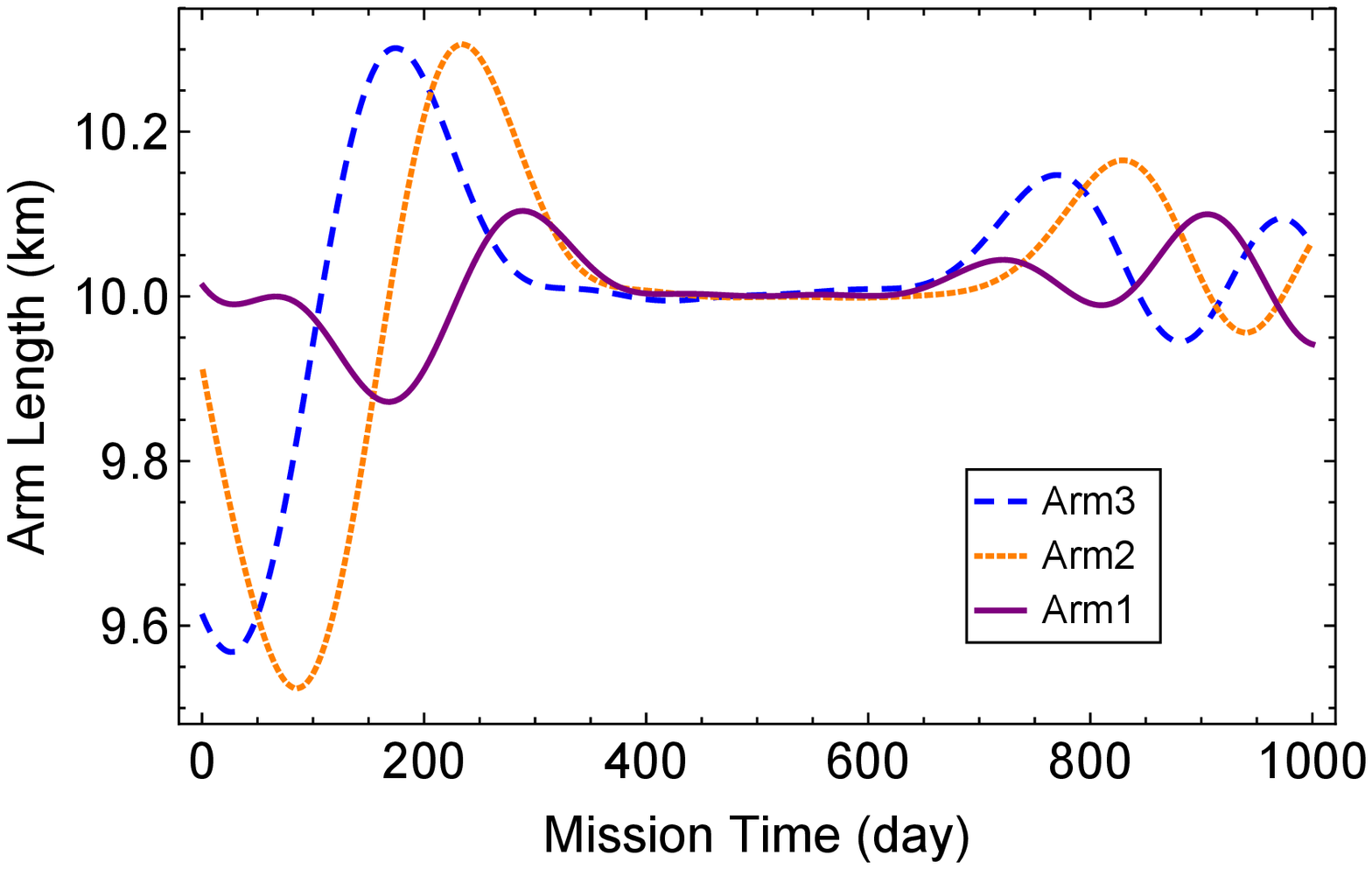}
   \includegraphics[width=0.49\textwidth]{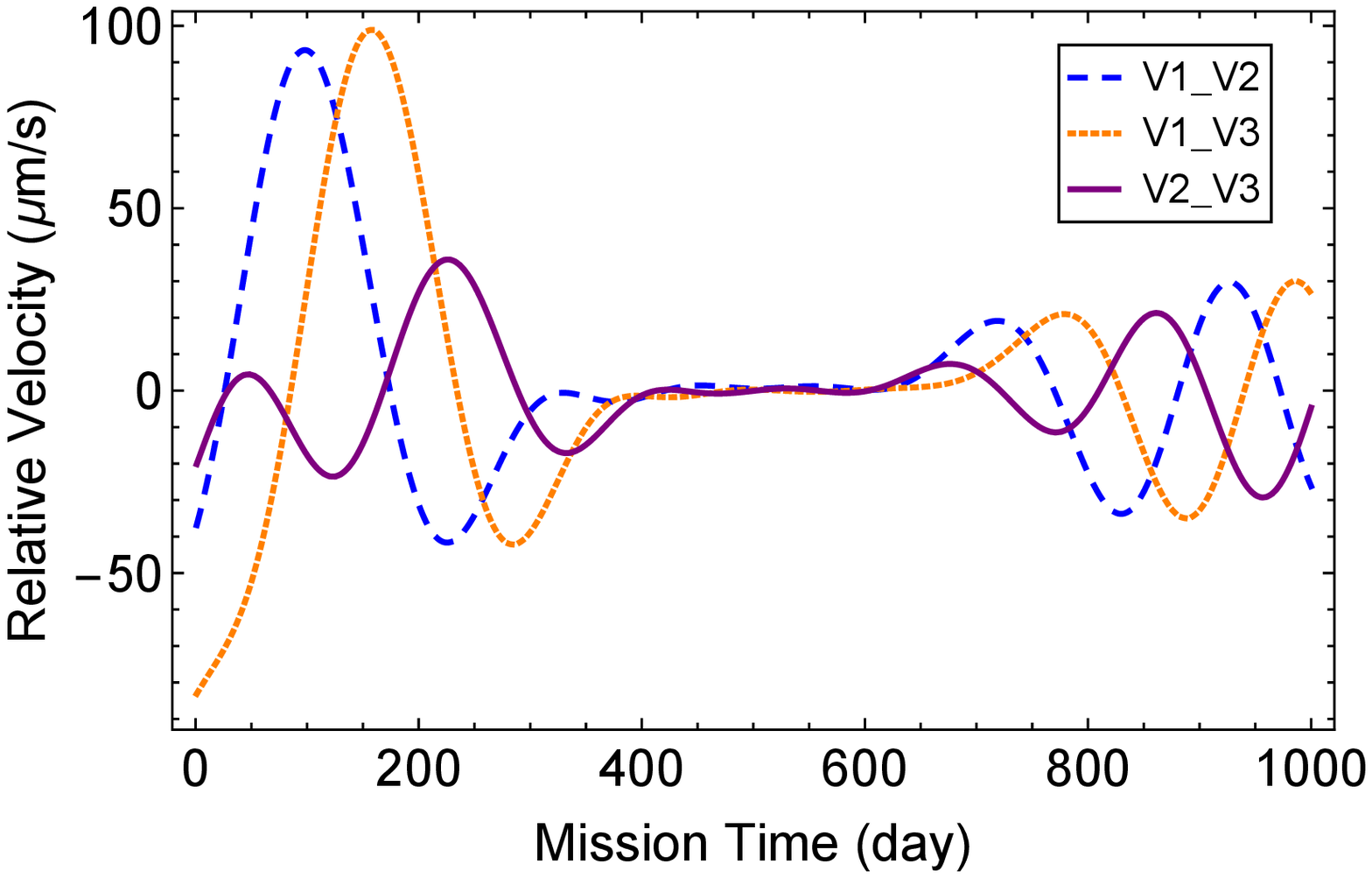} 
   \caption{The variation of arm lengths (left panel) and relative velocities in line-of-sight (right panel) for three arms of nominal 10-degree configuration in 1000 mission days.}
   \label{fig:3ams_to_pick}
\end{figure}

After we have the initial conditions from Eqs. (\ref{equ:1})-(\ref{equ:initialcondition}) for 10-degree trailing angle configuration, we run the 1000 days simulation from $t_0$ = JD2462503.0 (2030-Jan-1st 12:00:00) using our ephemeris framework CGC2.7.1\cite{wang&ni2015}, and find that the initial orbits have the relatively good equal-arm performance at the beginning, and become worse with the time expanding.
Then we empirically evolve the orbit backward in time for 500 days and reach JD2462003.0 (2028-Aug-19th 12:00:00). The orbit in this 500 days is comparable to the forward 500 days. Consequently, we combine these two periods result as shown in Fig. \ref{fig:3ams_to_pick}. As we can see in the Fig. \ref{fig:3ams_to_pick} left panel, the Arm1, formed by S/C2 and S/C3, is the most stable arm in the three. And the relative velocity between S/C2 and S/C3 also has smaller fluctuation than the other two pairs as shown in Fig. \ref{fig:3ams_to_pick} right panel. Therefore, we choose the S/C2 and S/C3 as two end S/Cs (S/CI and S/CIII in Fig. \ref{fig:AIGSO_SCs}) for AIGSO with 1000 mission days. For the middle S/C, it is named as S/CII.

\begin{figure}[htb]
   \centering
   \includegraphics[width=0.48\textwidth]{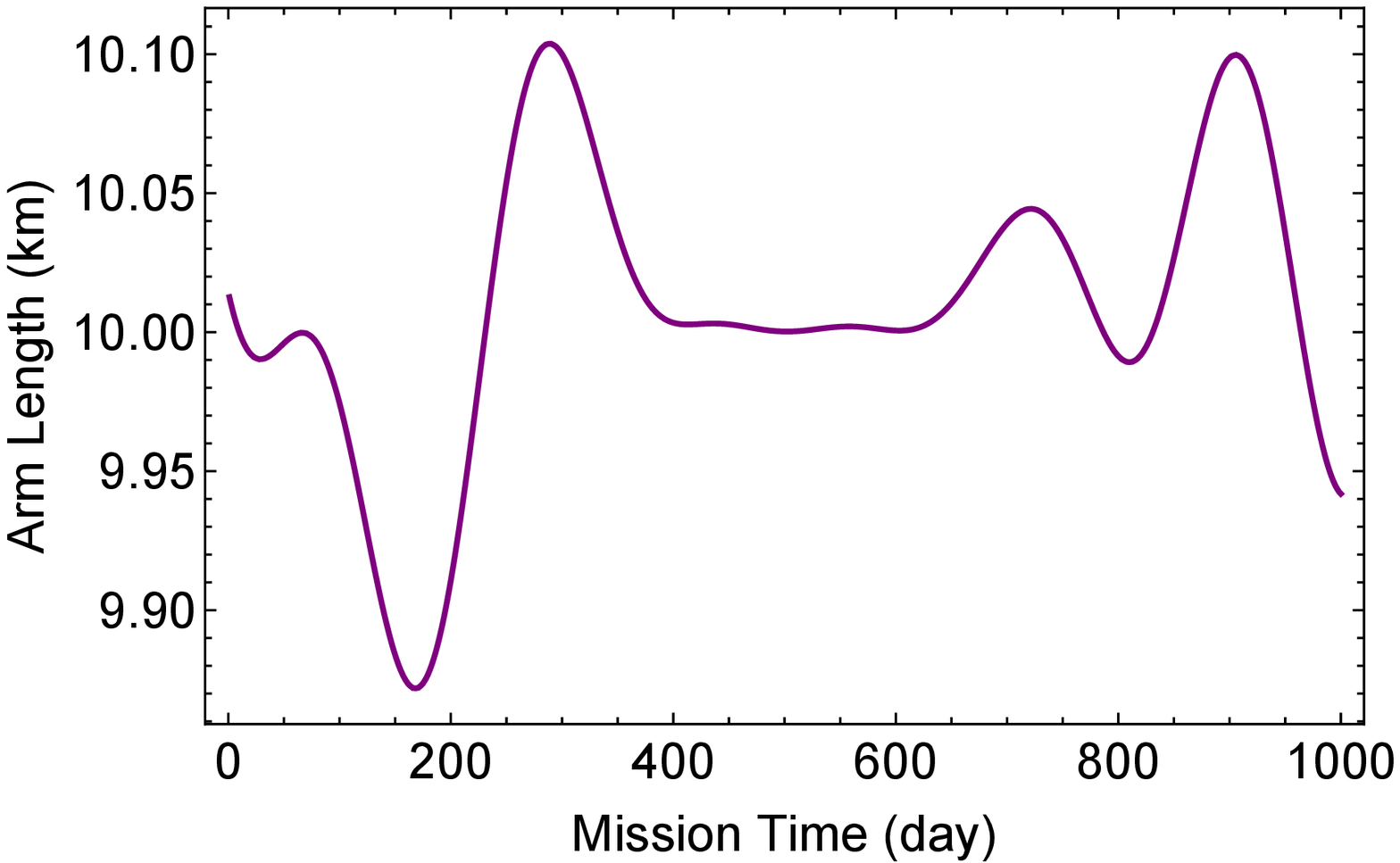} \ 
   \includegraphics[width=0.46\textwidth]{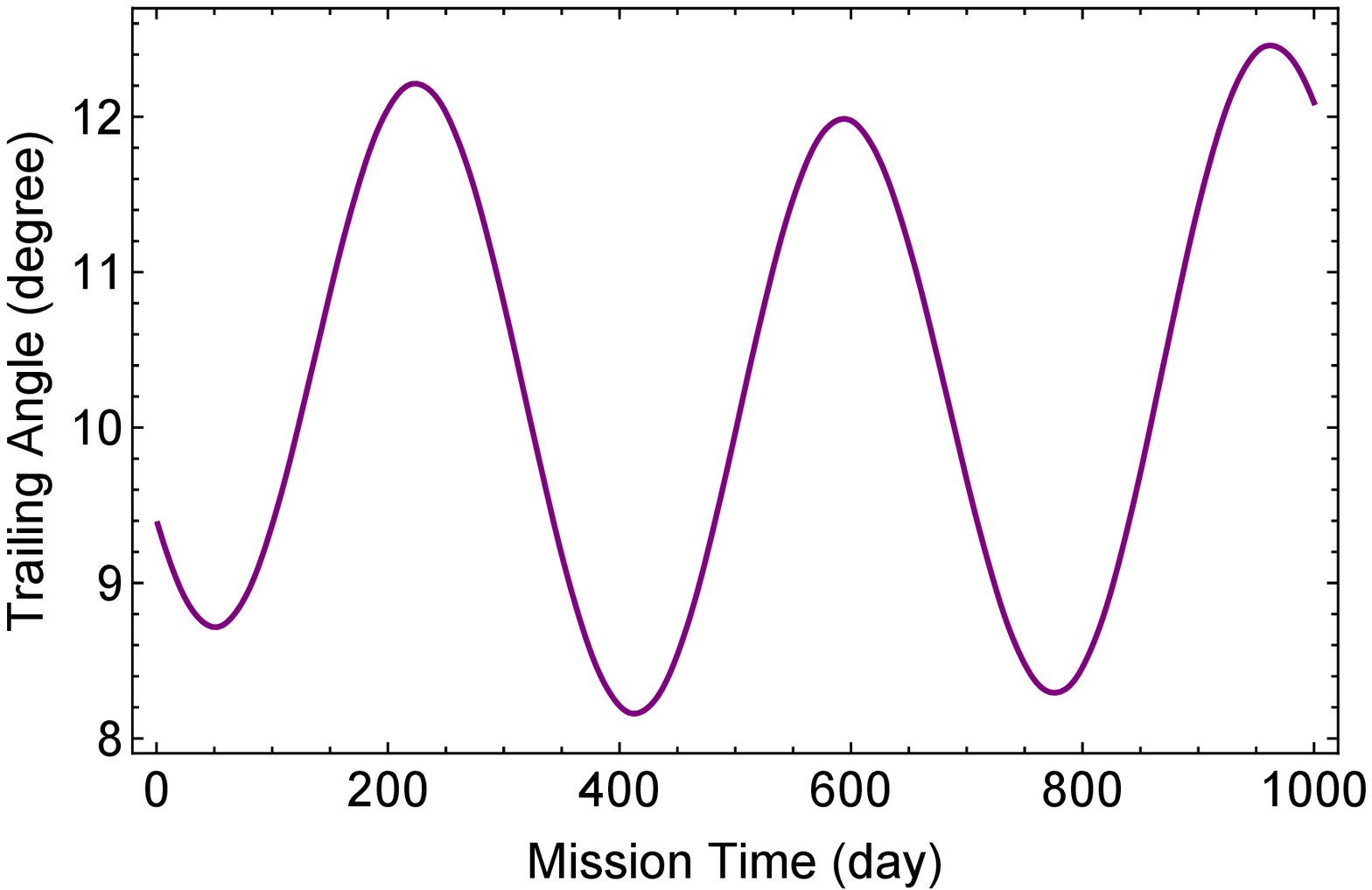} 
   \includegraphics[width=0.48\textwidth]{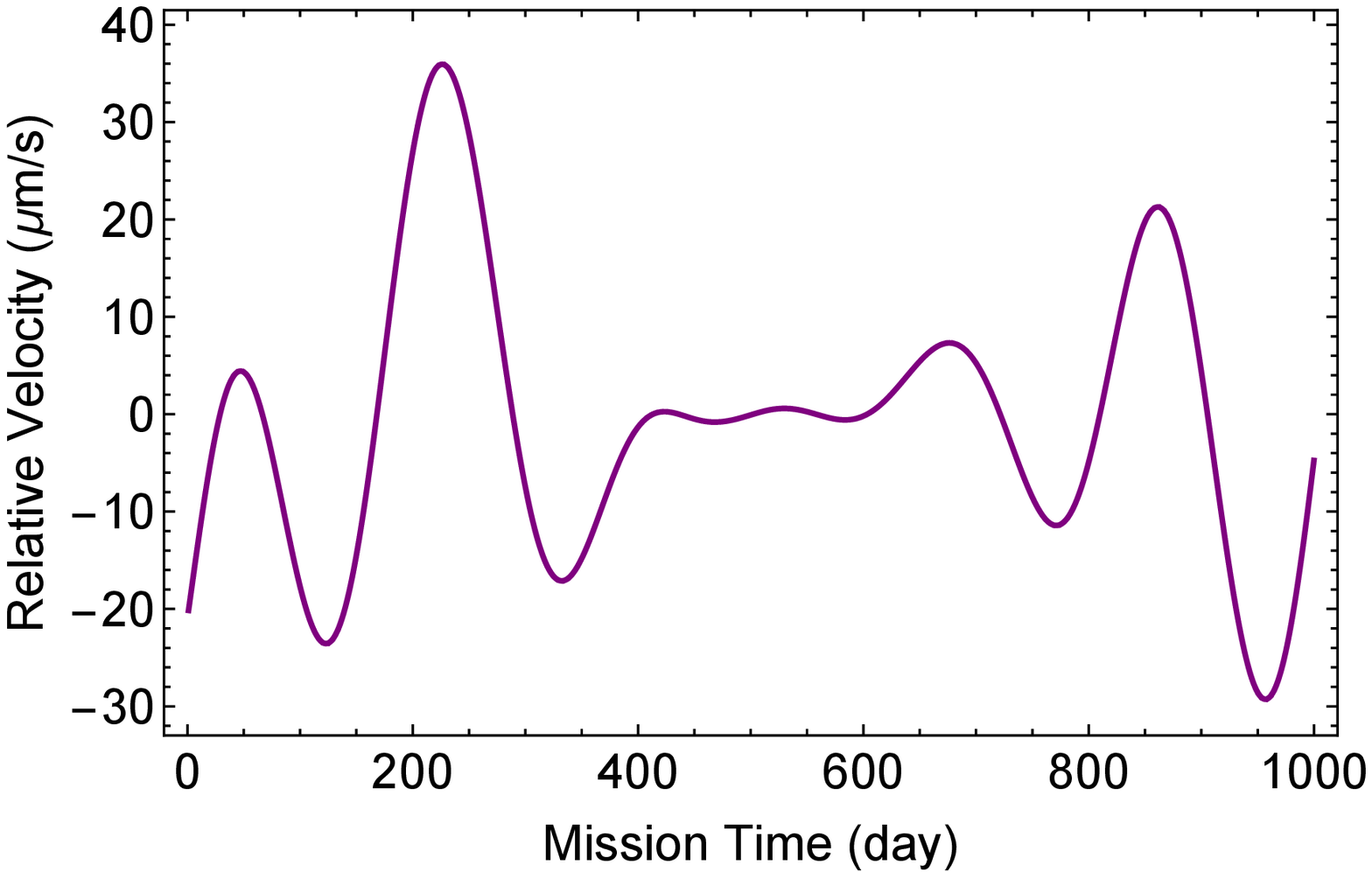} \ 
    \includegraphics[width=0.48\textwidth]{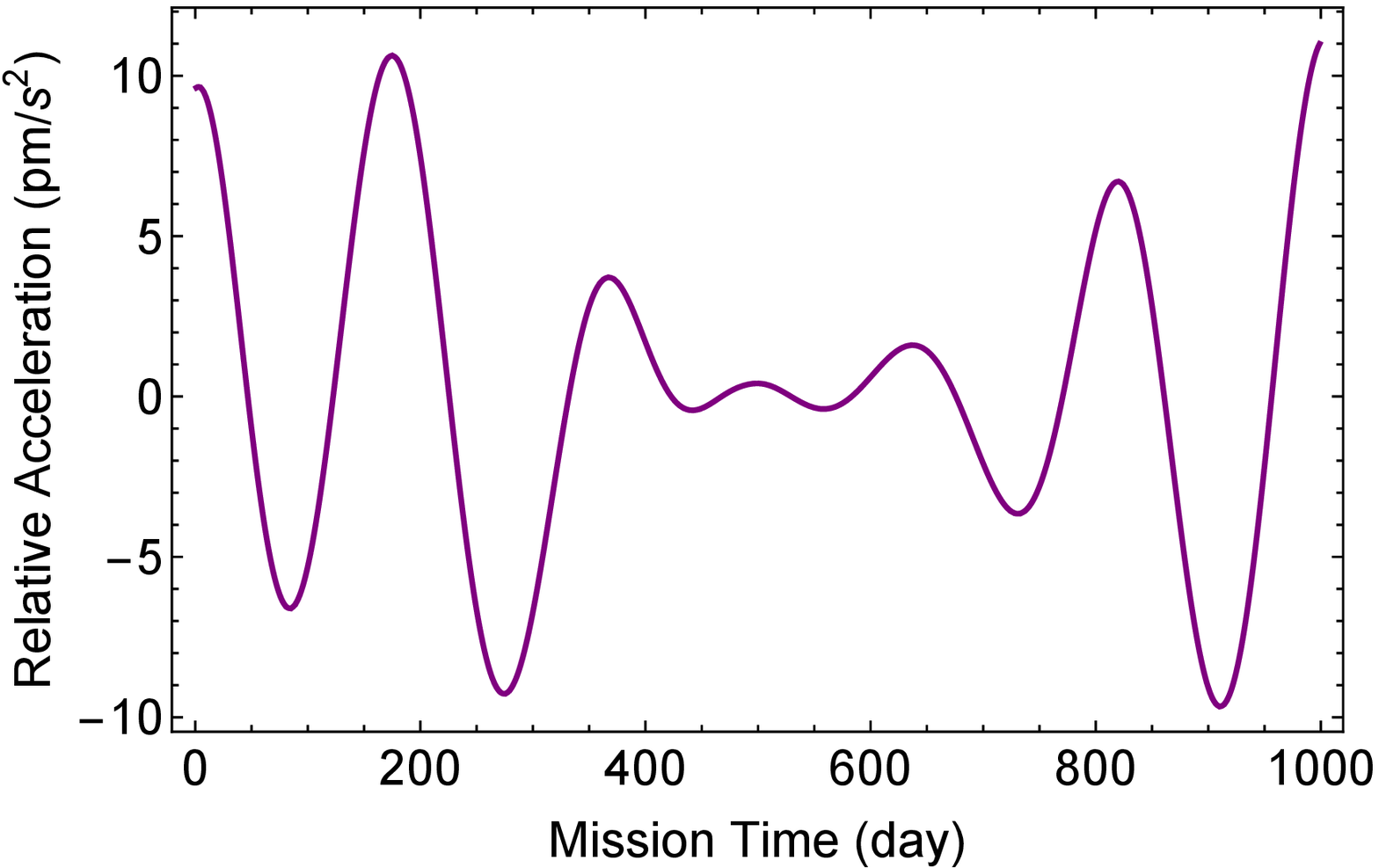}
    \includegraphics[width=0.48\textwidth]{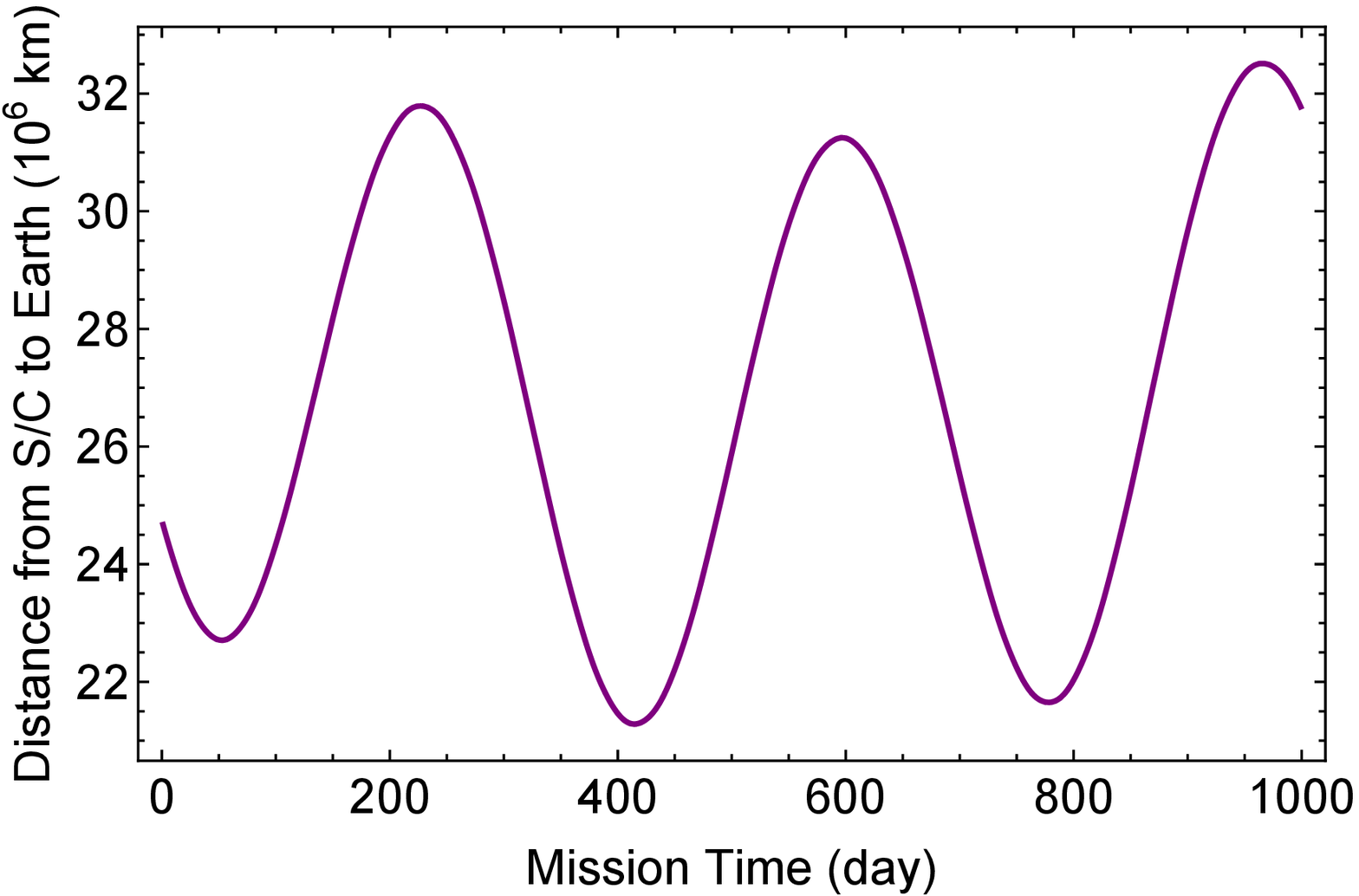}
   \caption{The variation of (a) arm lengths, (b) trailing angle, (c) relative velocities in line-of-sight, (d) relative acceleration in line of sight, (e) distance between Earth and S/C for the mission orbit of nominal 10-degree configuration.}
   \label{fig:orbit_10deg}
\end{figure}

The orbit variations with time for single-baseline AIGSO mission orbit are shown in Fig. \ref{fig:orbit_10deg}, which include arm length, trailing angle (angle behind the Earth), relative velocity (relative velocity in the line-of-sight), relative acceleration (relative acceleration in the line-of-sight) and total relative acceleration.
The fluctuation of the arm length is $\sim 0.1$ km which is $\sim 1\%$ of the arm length. The trailing angle changes in the range of $\sim[8^\circ,~12^\circ]$. The amplitude of relative velocity is smaller than $40~\mu$m/s, and the amplitude of relative acceleration is smaller than $12$ pm/s$^2$. The initial conditions of two S/Cs leading to this result are listed in Table \ref{tab:init_condition} third column.

\begin{figure}[htb]
   \centering
   \includegraphics[width=0.49\textwidth]{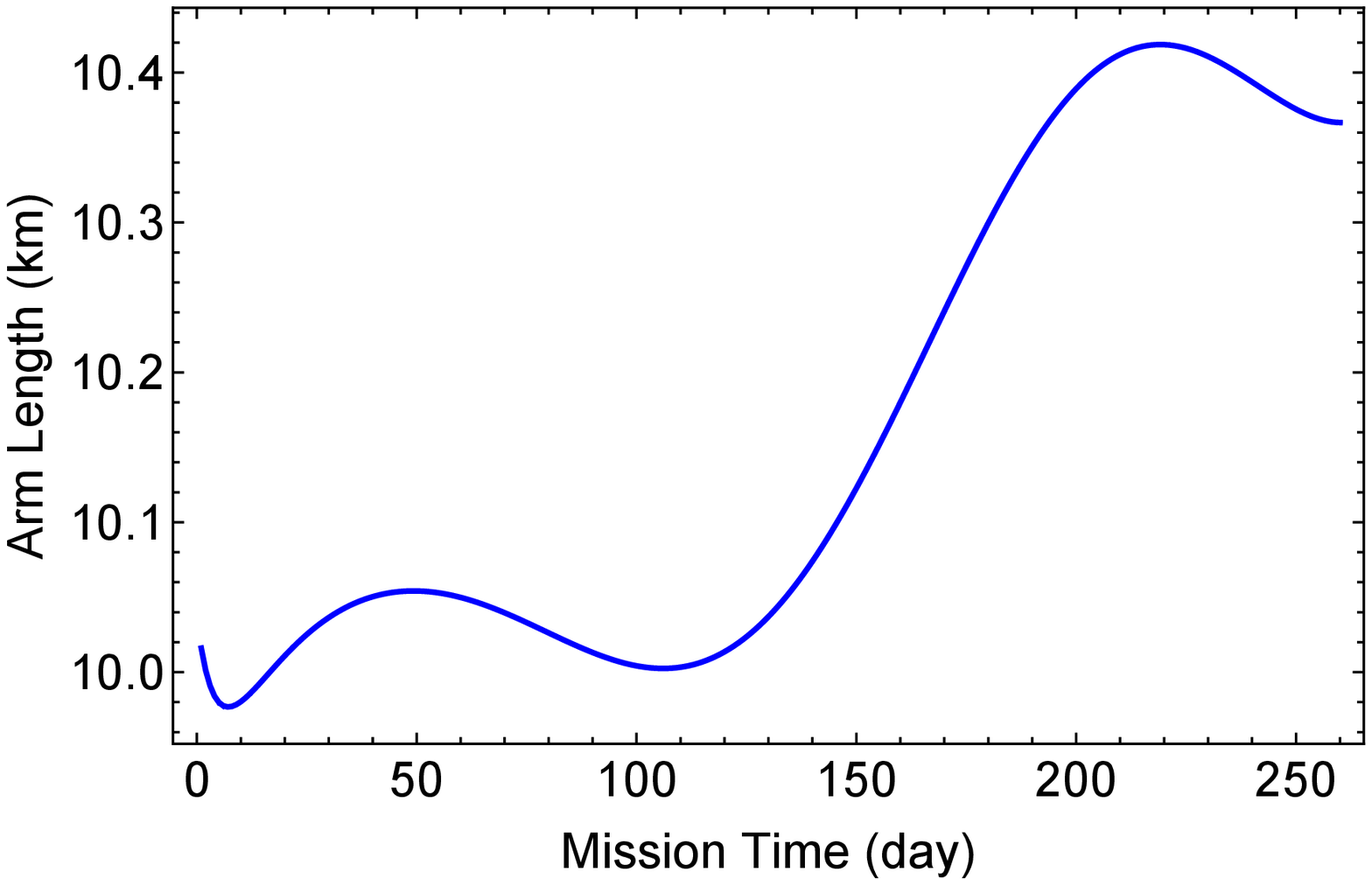} \ \ 
   \includegraphics[width=0.46\textwidth]{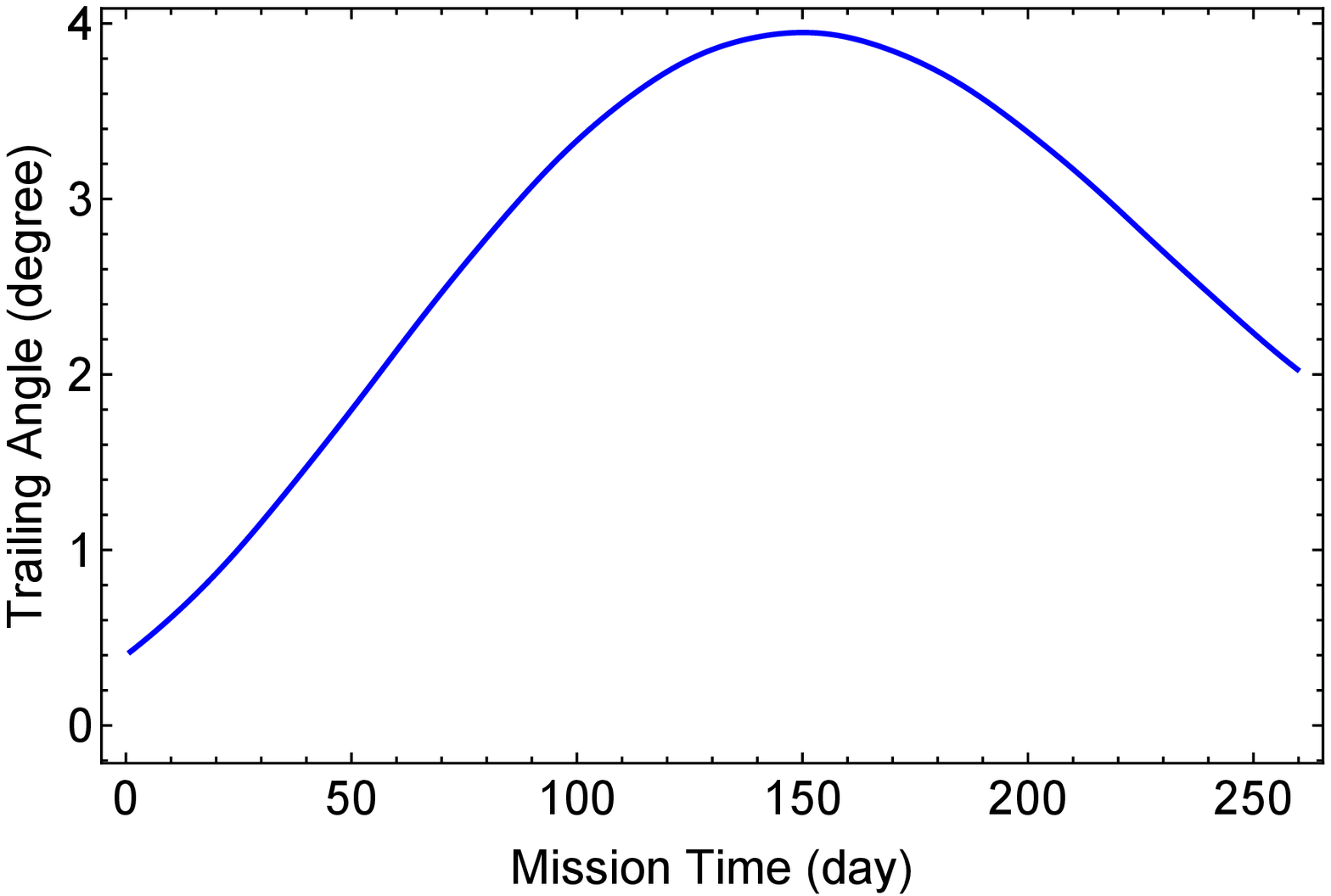}
   \includegraphics[width=0.49\textwidth]{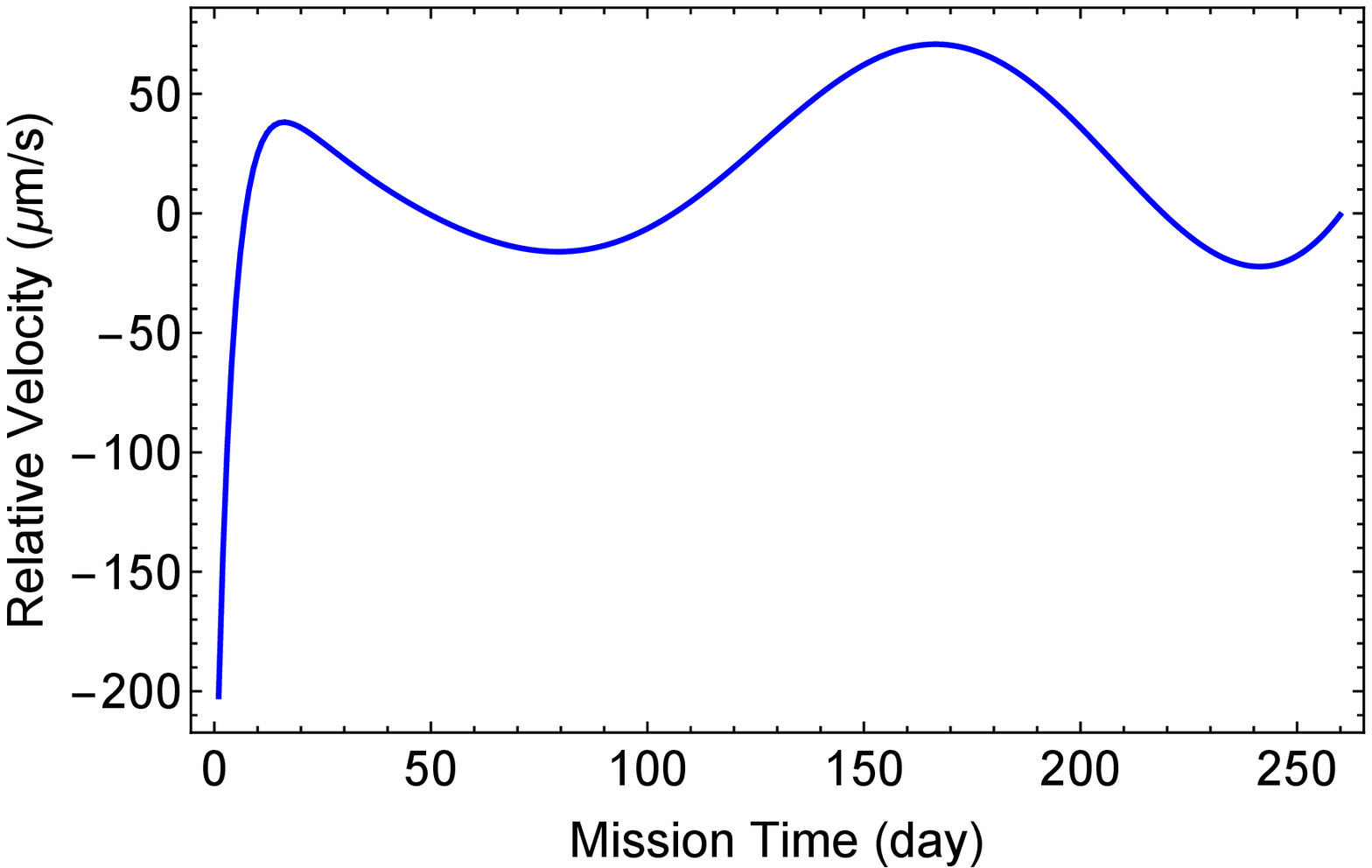} \ 
   \includegraphics[width=0.49\textwidth]{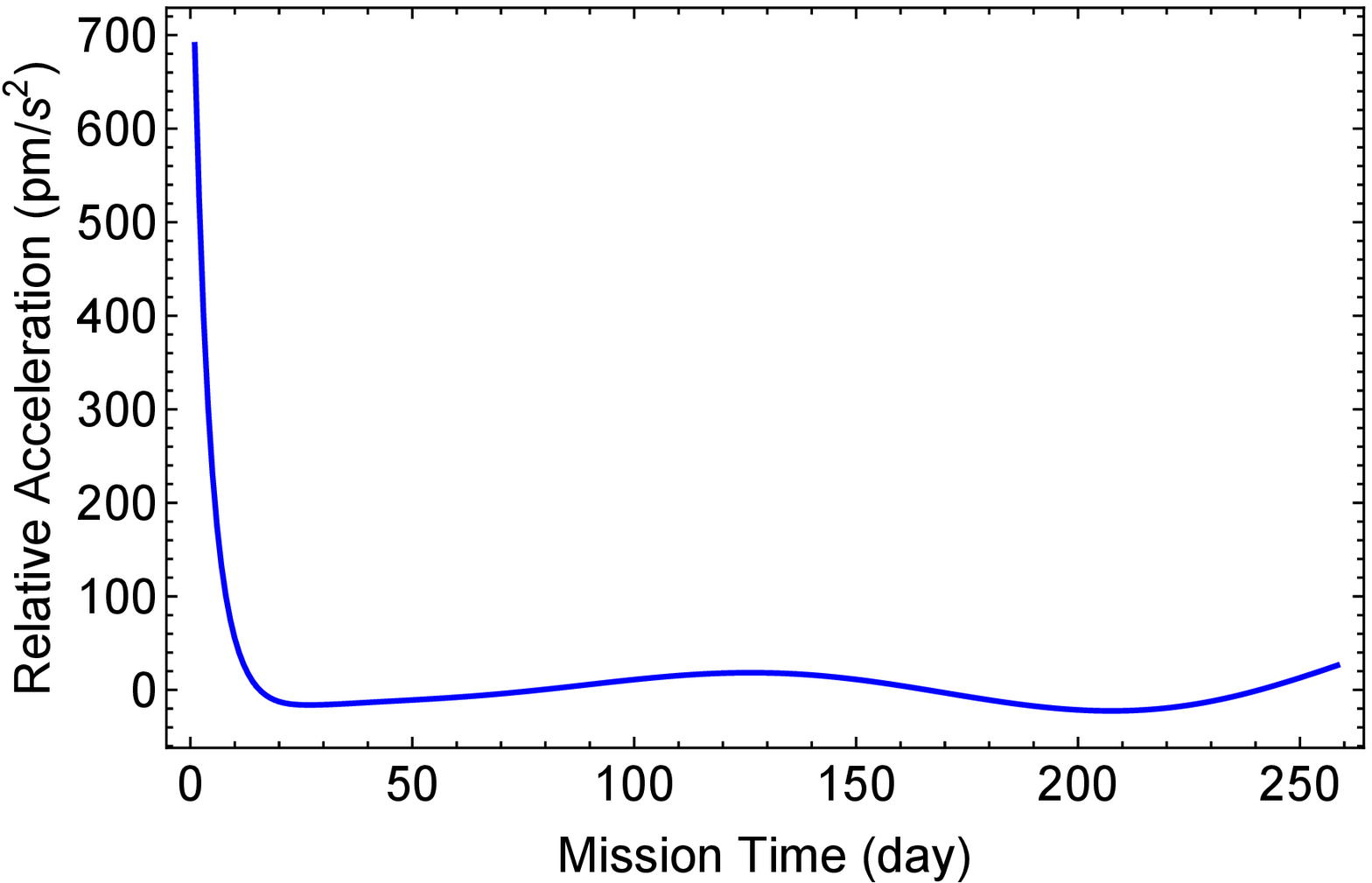}
   \includegraphics[width=0.49\textwidth]{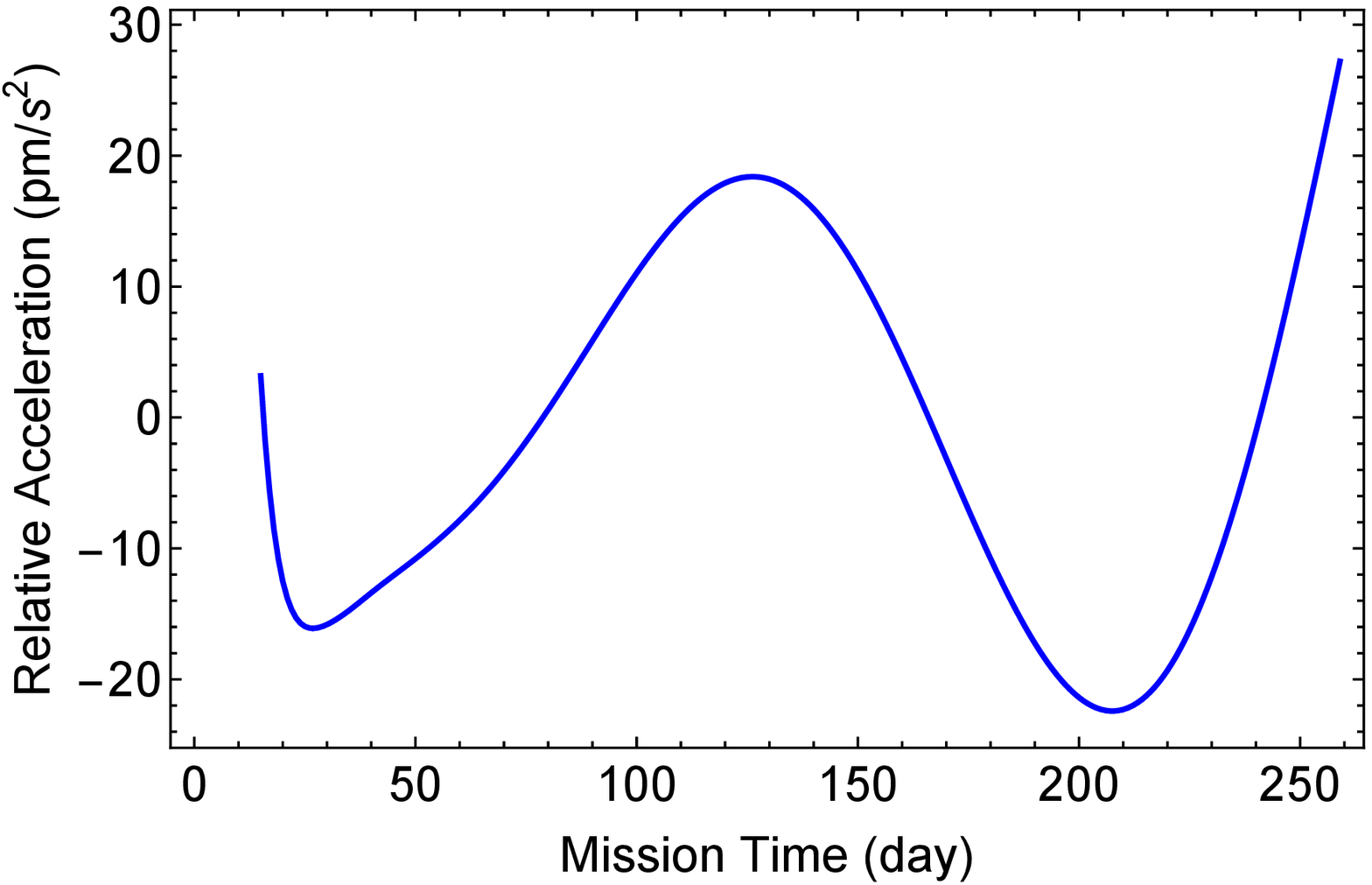} \ 
   \includegraphics[width=0.49\textwidth]{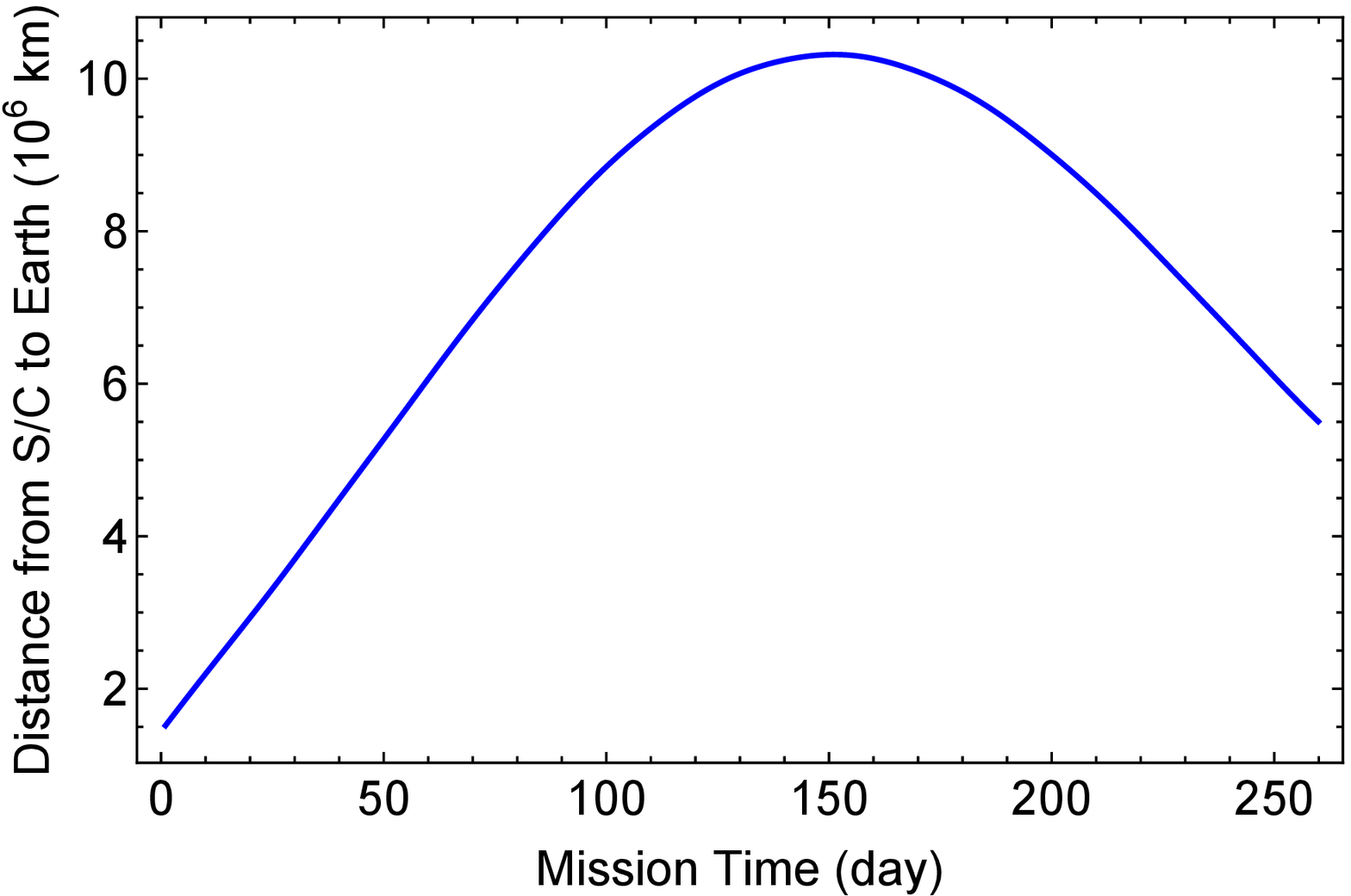}
   \caption{The variation of (a) arm lengths, (b) trailing angle, (c) relative velocities in line-of-sight, (d) relative acceleration in line of sight, (e) relative acceleration in line of sight after truncating the first 15 days and (f) distance between Earth and S/C for the mission orbit of nominal 2-degree configuration.}
   %(f) total relative acceleration after truncating the first 15 days
   \label{fig:orbit_2deg}
\end{figure}

For the orbit with 10-degree trailing angle, the S/Cs is $2.6 \times 10^{7}$ km away from Earth, the orbit could be in a stable status for at least 1000 days as we simulated. While the mission cost may also be higher than the mission which is closer to the Earth. To explore the possible qualified orbit closer to Earth, using the method above, we initially choose the 2-degree trailing angle at the same epoch $t_0 =$ JD2462503.0 (2030-Jan-1st 12:00:00). Then we evolve the orbit forward and backward as we did previously. We select the time JD2462448.0 (2029-Nov-07 12:00:00) as our new orbit starting time and simulate the 260 days orbit as shown in Fig. \ref{fig:orbit_2deg}. 

Due to the Earth and Moon's gravitational perturbations, the stability of arm length becomes weaker than the orbit with 10-degree lagging angle. We only demonstrate the 260 days result. As we can see from the plots in Fig. \ref{fig:orbit_2deg}, the variation of arm length in 260 days is in 5\%. The trailing angle is 0.5 deg at the initial position which the geocentric distance is $1.3 \times 10^{6}$ km, and it increases and reaches the 4 degrees at 150 days (nonetheless, we keep labeling this configuration as nominal 2-degree trailing angle). While the Sun-Earth Lagrangian point L1/L2 geocentric distance is $\sim 1.5 \times 10^{6}$ km.
The velocities and accelerations in line-of-sight direction in the first 15 days change rapidly as shown in the middle row of Fig. \ref{fig:orbit_2deg}. After 15 days, the mission orbit becomes relatively stable as we plotted in Fig. \ref{fig:orbit_2deg} last row left panel. The initial conditions of the two S/Cs in the J2000 equatorial solar-system-barycentric coordinate system are listed in Table \ref{tab:init_condition} fourth column. 

\begin{table}[ht]
\tbl{The initial conditions of the three AIGSO orbit configurations (10-degree and 2-degree trailing angle with inclination together with 10-degree with no inclination) in J2000 equatorial solar-system-barycentric coordinate system.}
{\begin{tabular}{@{}ccccc@{}} \toprule
& & 10-degree & 2-degree & 10-degree trailing \\
 &  & trailing angle & trailing angle &   angle in ecliptic plane  \\
& & (JD2462003.0) & (JD2462448.0) & (JD2462343.0) \\ \colrule
 S/C2       & $X$ & 7.344240676740E-01 & 7.122030884852E-01 & 3.909036563191E-01  \\  
 Position & $ Y$ & -6.237915670030E-01 & 6.421941165460E-01 & -8.445799436780E-01  \\
  (AU)      & $Z$ & -2.704122467620E-01 & 2.784585667244E-01 &  -3.661360099923E-01 \\
  S/C2     & $V_x$ & 1.169136940650E-02 & -1.192677883933E-02 & 1.584402750434E-02   \\
Velocity &$V_y $ & 1.159542794680E-02 & 1.129944459347E-02  & 6.156504193175E-03 \\
   (AU/day)  &$V_z$ & 5.027215158430E-03 & 4.899412462931E-03 & 2.669194583507E-03 \\
\colrule
S/C3 & $X$ & 7.344241170660E-01 & 7.122031379384E-01 & 3.909035953722E-01   \\
Position & $Y$ & -6.237915680550E-01 & 6.421940766354E-01 & -8.445799679030E-01 \\
(AU) & $Z$ & -2.704122016070E-01 & 2.784585456477E-01 & -3.661360204960E-01 \\
S/C3 & $V_x$ & 1.169136882050E-02 & -1.192677910407E-02 & 1.584402794360E-02 \\
Velocity & $V_y$ & 1.159542763950E-02 & 1.129944500437E-02 &  6.156503232243E-03 \\
(AU/day) & $V_z$ & 5.027215775010E-03 & 4.899411434559E-03 &  2.669194166894E-03 \\
\colrule
\end{tabular} \label{tab:init_condition}}
\end{table}

\section{Orbit Correction and Thruster Requirement} \label{sec:fixed-baseline}

From the geodesic orbit achieved in Section \ref{sec:geodesic}, the arm length of AIGSO changes with time. To retain the constant baseline arm length, the thrusters are needed to adjust the position of the S/C. In this section, we investigate the thruster requirement in AIGSO mission.

\subsection{Thruster Acceleration}

As shown in Fig. \ref{fig:AIGSO_SCs}, the S/CI is the spacecraft to send out the atomic beams. If the geodesic orbit of S/CI is chosen as the fiducial orbit, the required trajectories of S/CII and S/CIII to keep the baseline could be calculated by
\begin{equation} \label{equ:traj_position}
\begin{split}
 \mathbf{r}_{\mathrm{traj, S/CI}} &= \mathbf{r}_{\mathrm{S/C}2} \\
 \mathbf{r}_{\mathrm{traj, S/CII}} &= \mathbf{r}_{\mathrm{S/C}2} + \frac{ \mathbf{r}_{\mathrm{S/C}3} -  \mathbf{r}_{\mathrm{S/C}2}}{ | \mathbf{r}_{\mathrm{S/C}3} -  \mathbf{r}_{\mathrm{S/C}2} | }  \times l/2, \\
 \mathbf{r}_{\mathrm{traj, S/CIII}} &= \mathbf{r}_{\mathrm{S/C}2} + \frac{ \mathbf{r}_{\mathrm{S/C}3} -  \mathbf{r}_{\mathrm{S/C}2}}{ | \mathbf{r}_{\mathrm{S/C}3} -  \mathbf{r}_{\mathrm{S/C}2} | } \times l \\
 \end{split}
\end{equation}
where $l=10$ km is the AIGSO proposed baseline length, $\mathbf{r}_{\mathrm{S/C}2,3} $ is the geodesic position of S/C$2,3$ calculated in Section \ref{sec:geodesic}. From Eq. (\ref{equ:traj_position}) we can obtain the acceleration at a specific point in a trajectory by calculating the second derivative of position w.r.t. time.
\begin{equation} \label{equ:traj_acceleration}
 \mathbf{a}_{\mathrm{traj}} = \mathbf{\ddot{r}}_{\mathrm{traj}}.
\end{equation}
The $ \mathbf{a}_{\mathrm{traj}}$ is the acceleration to follow a trajectory. However, since the trajectories from Eq. (\ref{equ:traj_position}) are not geodesics, the accelerations from Eq. (\ref{equ:traj_acceleration}) should be different from the gravitational acceleration in the solar system. The deviations between these two kinds of acceleration should be provided by the thruster action.

To obtain the instantaneous gravitational acceleration of S/CII and S/CIII in the solar system at a specific world point (i.e., a specific 3-dimension point at a certain epoch) in the trajectory, we insert the positions of S/CII and S/CIII from Eq. (\ref{equ:traj_position}) into our ephemeris frame CGC2.7.1. The accelerations considered in our ephemeris frame include: 1). the Newtonian and post-Newtonian gravitational acceleration from the Sun, 8 major planets, Pluto, Moon, Ceres, Pallas, and Vesta\cite{Brumberg1991}, 2) the acceleration due to the figure effects from the Sun and Earth (the Sun 2nd-degree and the Earth 2-4 degree zonal harmonics), 3) the Newtonian perturbations from the selected 340 asteroids\cite{Folkner2014,asteroid},
\begin{equation} \label{equ:a_eph}
  \mathbf{a}_\mathrm{eph} ( \mathbf{r}_{\mathrm{traj}} , \mathbf{\dot{r}}_{\mathrm{traj}})= \mathbf{a}_{\mathrm{Newton}} + \mathbf{a}_{\mathrm{1PN}} +  \mathbf{a}_{ \mathrm{fig}} +  \mathbf{a}_{\mathrm{asteroid}}.
\end{equation}
The explicit form of Eq. (\ref{equ:a_eph}) of interactions in the CGC ephemeris framework is fully described in references \refcite{wang&ni2012,wang2011,wang&ni2015}.

\begin{figure}[htb]
   \centering
   \includegraphics[width=0.49\textwidth]{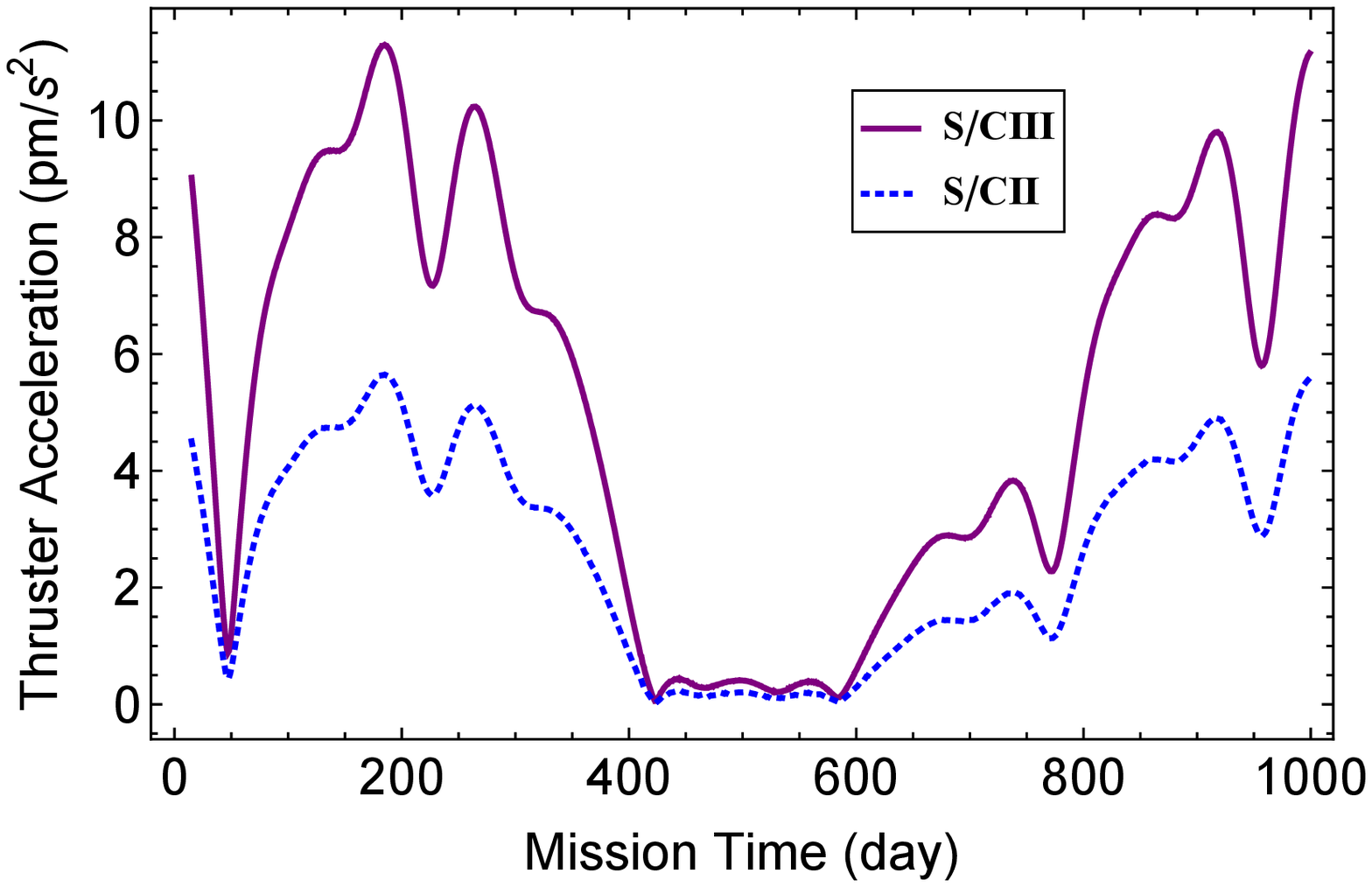} \ 
   \includegraphics[width=0.48\textwidth]{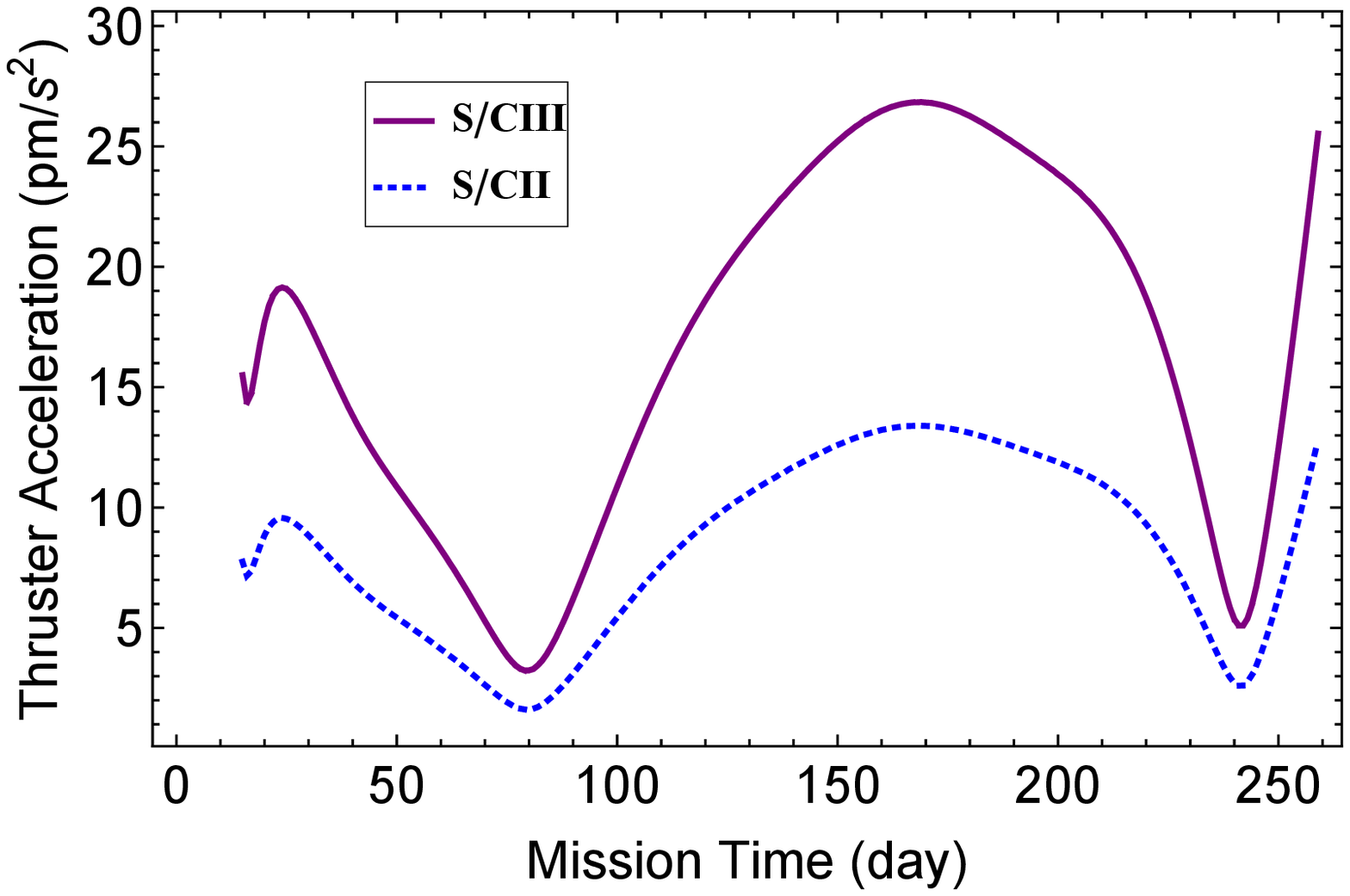} 
   \caption{The thruster acceleration compensations to maintain the AIGSO baseline arm length for S/CII and S/CIII in the ephemeris framework CGC2.7.1 for 10-degree (left panel) and 2-degree (right panel) trailing angle configurations, respectively.}
   \label{fig:diff_A}
\end{figure}

The thruster acceleration to maintain the constant arm length trajectories is calculated by
\begin{equation}
 \mathbf{a}_{\mathrm{thruster}} =  \mathbf{a}_{\mathrm{traj}} - \mathbf{a}_\mathrm{eph}.
\end{equation}
The results for 10-degrees and 2-degree lagging angle configurations are shown in Fig. \ref{fig:diff_A}. It is less than $\sim 30$ pm/s$^2$ for the designed trajectories of S/CII and S/CIII. Assuming the weight of a S/C is 1000 kg, thrusters with 30 nano-Newton (30 nN) is required to adjust the S/C. This requirement is ready met by the current thruster technology.

\subsection{Alternate Linear Formation and Its Thruster Acceleration Compensation} \label{sec:4.2}

\begin{figure}[htb]
   \centering
   \includegraphics[width=0.48\textwidth]{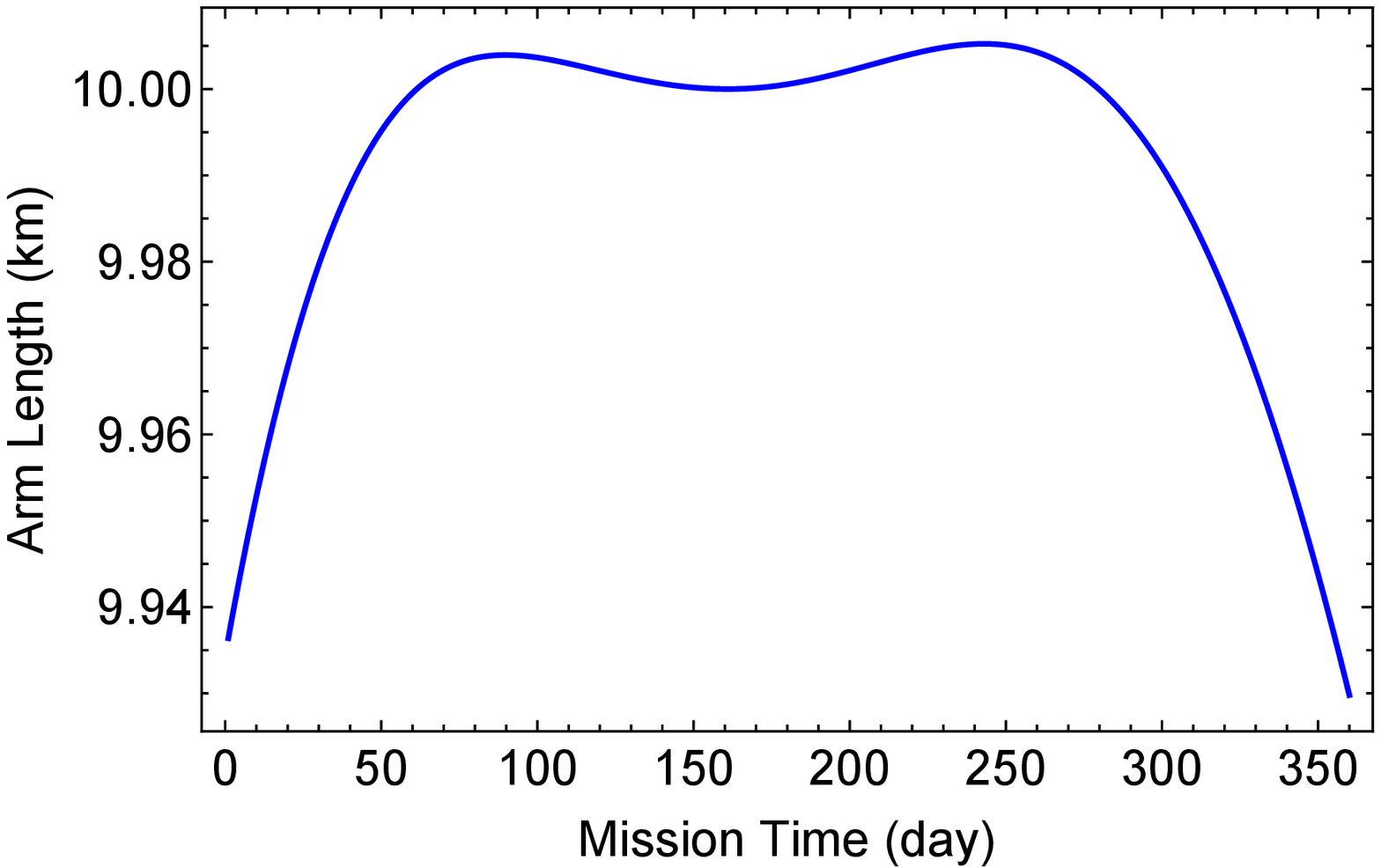} \ 
   \includegraphics[width=0.47\textwidth]{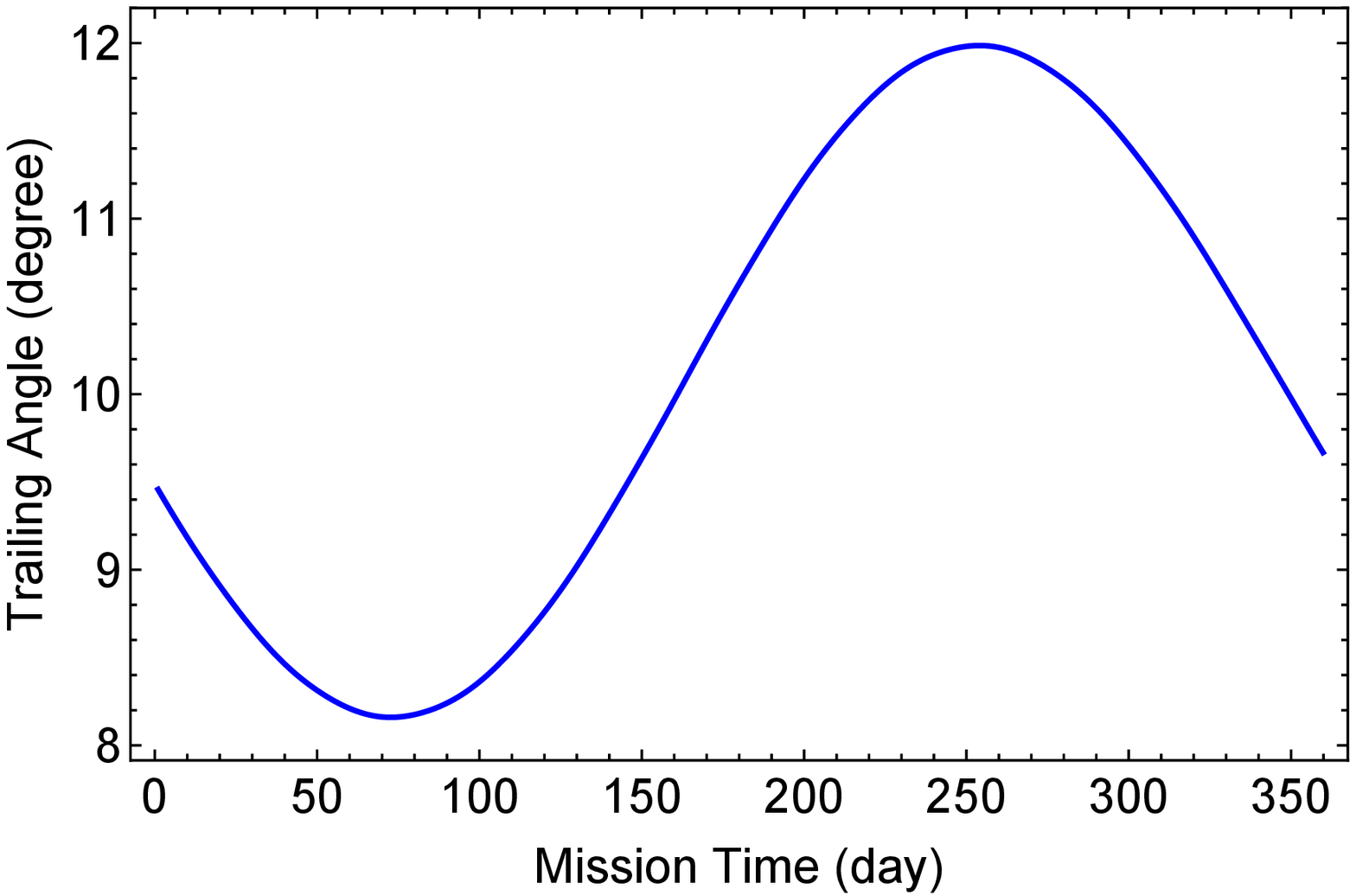}
   \includegraphics[width=0.48\textwidth]{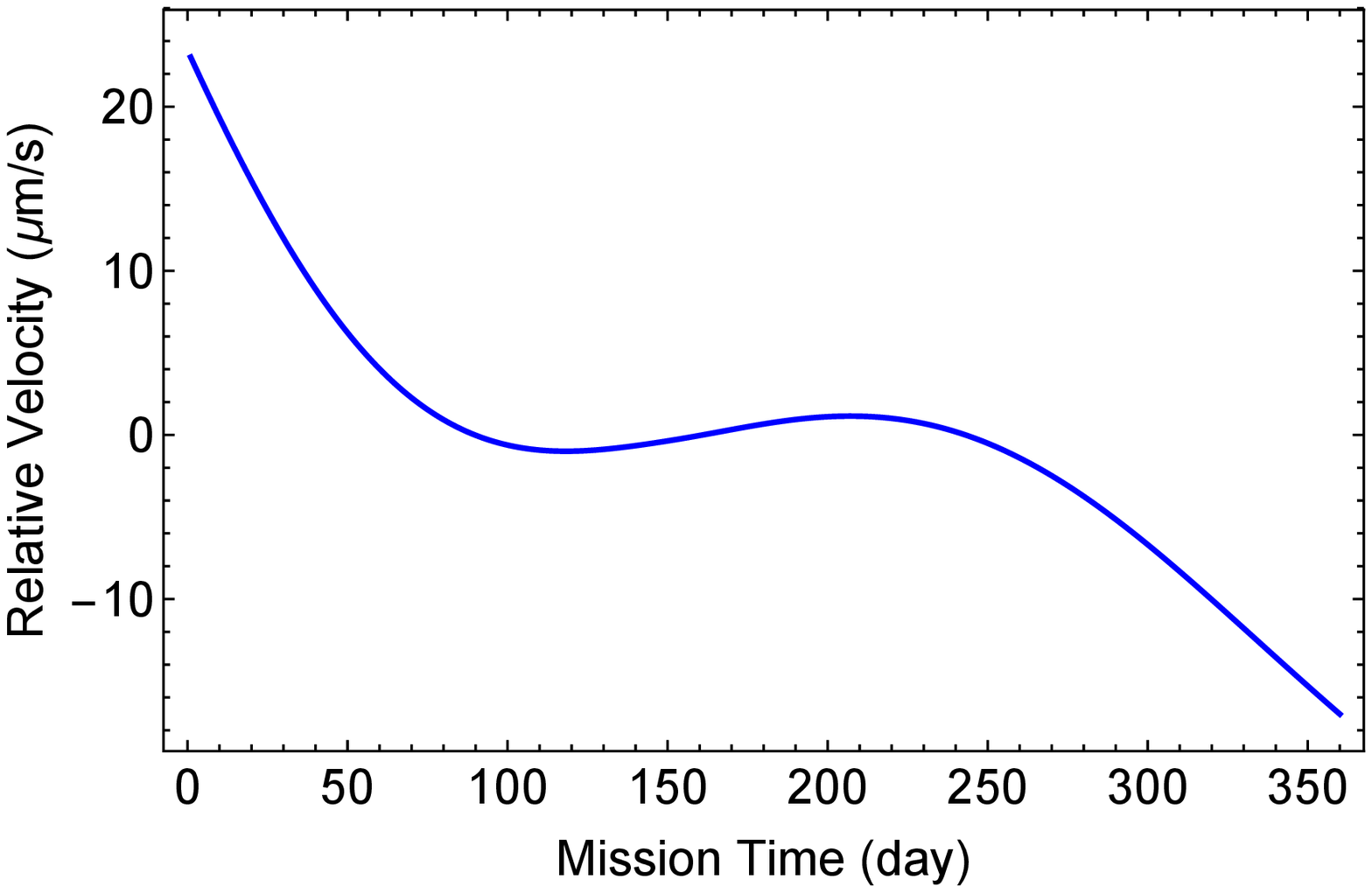} \ 
   \includegraphics[width=0.48\textwidth]{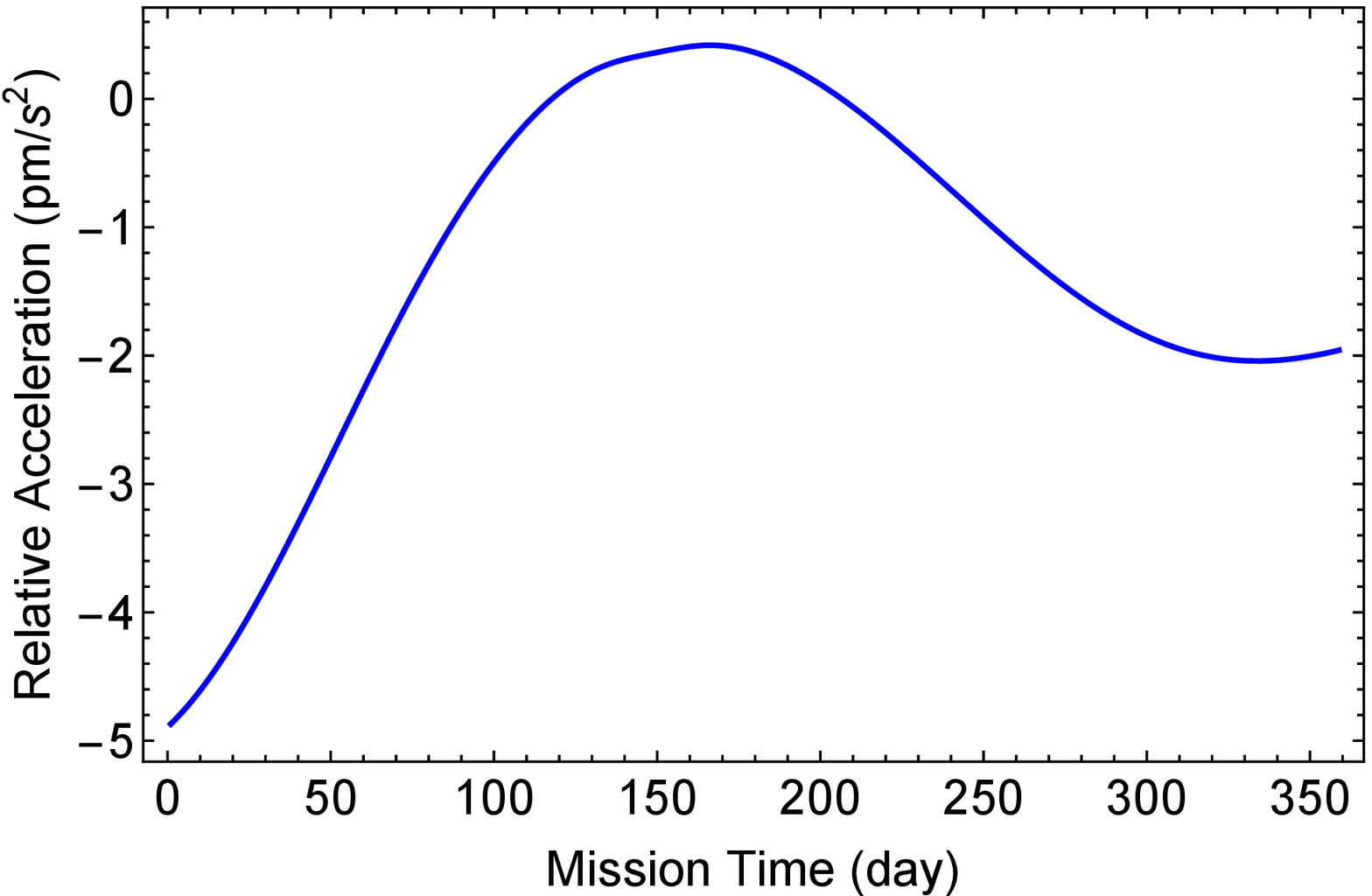}
   \includegraphics[width=0.48\textwidth]{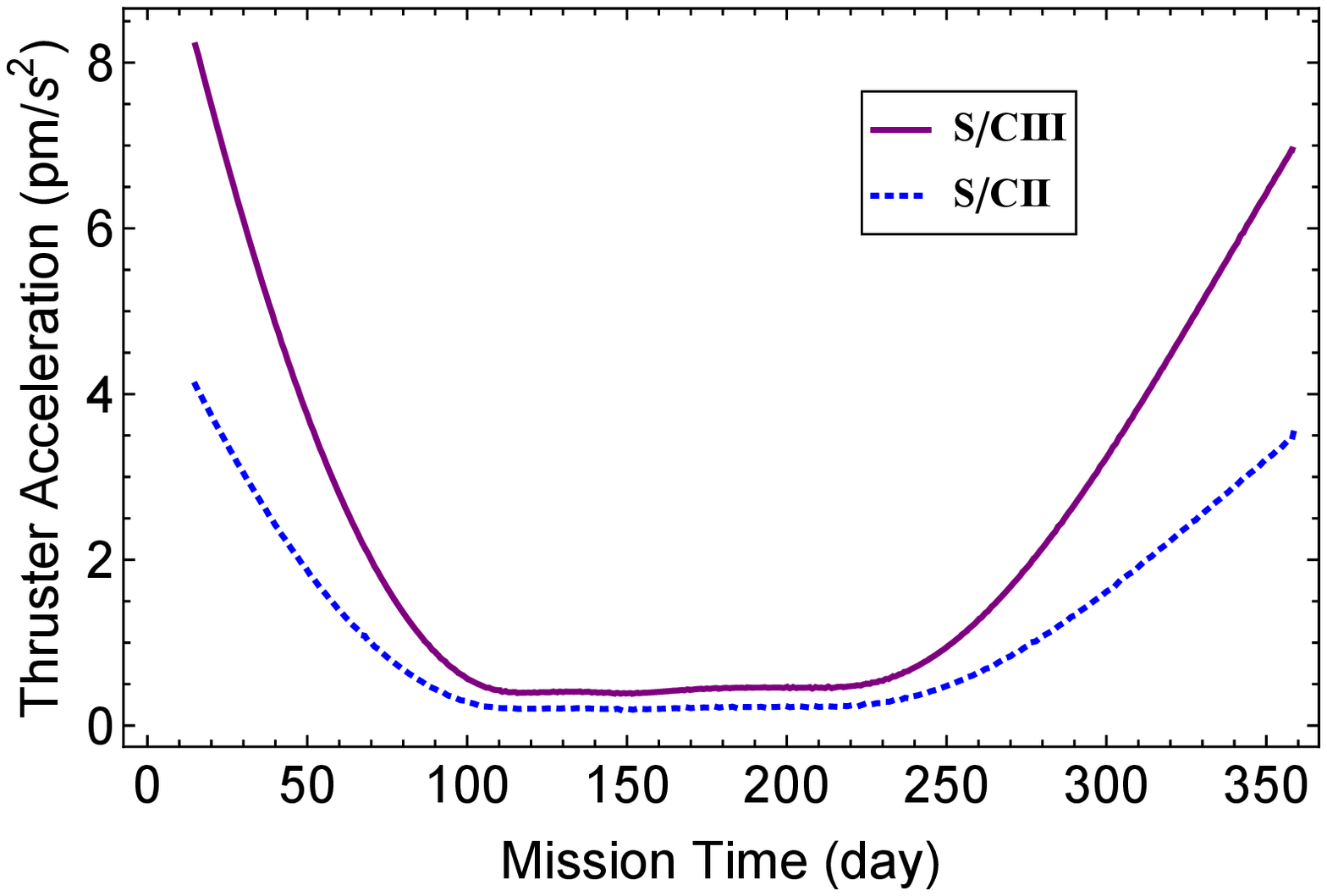} \ 
   \caption{The variation of (a) arm lengths, (b) trailing angle, (c) relative velocities in line-of-sight, (d) relative acceleration in line of sight, (e) thruster acceleration requirement to keep the baseline for a mission orbit of in ecliptic plane with nominal 10-degree trailing angle.}
   \label{fig:orbit_10deg_inplane}
\end{figure}

In Section \ref{sec:AIGSO}, we applied CW frame to design a nearly constant formation. This approach is necessary for 2-dimensional formation. However, if AIGSO needs only 1-dimensional formation, this can be in the ecliptic plane for AIGSO mission concept. For this alternative, we simulate with an orbit around the Sun in the ecliptic plane with 10-degree trailing angle. The plots of 360 days mission orbit without optimization are shown in Fig. \ref{fig:orbit_10deg_inplane}. The arm length variations are comparable to the result we have achieved in Section \ref{sec:geodesic}. the thruster requirement is also at the same level. The initial conditions of the S/Cs are listed in the fifth column of Table \ref{tab:init_condition}.

\section{Discussions and Conclusions} \label{sec:conclusion}

(i) The linear orbit configuration with $\sim60$ degrees inclination with constant arm length of AIGSO could be maintained with thruster requirement $\sim 15$ pm/s$^2$ in 3 years with lagging behind the Earth angle change from 8 degrees to 12 degrees, and with $\sim 30$ pm/s$^2$ in 245 days with lagging behind the Earth angle range from 0.5 degree to 4 degrees, respectively. The case in the ecliptic plane without inclination of Section \ref{sec:4.2} also satisfies a similar thruster requirement.

(ii) The 0.5 degree to 4 degrees case could be used as a pathfinder formation orbit. The transfer orbit for this case requires shorter period from low Earth orbit (200 km altitude) as in the case of AMIGO. This orbit could also be used as the first part of the transfer orbit for the 8 degrees to 12 degrees case.

(iii) In actual feedback of thrusters to maintain the constant formation, the thruster action and the inertial mass position adjustment could be done in alternate time lapse. How this would be implemented needs careful studies.

(iv) The method used in this work could also be applied to other mission concepts which need to maintain the orbit configuration precisely, for instance, B-DECIGO and DECIGO.

\section*{Acknowledgments}
This work was supported by the National Key Research and Development Program of China under Grant No. 2016YFA0302002, and the Strategic Priority Research Program of the Chinese Academy of Sciences under Grant No. XDB21010100.

%\appendix
%\begin{thebibliography}{000} %for 3 digits
%\begin{thebibliography}{00}  %for 2 digits

\end{document}